\documentclass[12pt]{article}
\usepackage[english]{babel}
\usepackage{cite}
\usepackage{amsmath,amssymb}
\usepackage{epsfig}

\begin{document}
\thispagestyle{empty}

\vspace*{2.0cm}

\begin{center}
\boldmath
 {\large \bf
Qualitative analysis of proton inelastic scattering for diquark searching
 }
\unboldmath
\end{center}
 \vspace*{0.8cm}

\begin{center}
{\sc Vladimir~V.~Bytev,$^{a}$\footnote{E-mail: bvv@jinr.ru}
Stepan.~S.~Shimanskiy$^{a}$ } \\
 \vspace*{1.0cm}
{\normalsize $^{a}$ Joint Institute for Nuclear Research,} \\
{\normalsize $141980$ Dubna (Moscow Region), Russia}
\end{center}

\begin{abstract}
 In this paper we discuss  exclusive reactions which analysis can be used to receive direct indication of diquark existence. We make estimations of  diquark scattering process measurement in inelastic proton-proton collisions. It was shown that putting special restrictions over kinematics and particles in final state of process  it will be possible to enhance potential diquark contribution to scattering  up to $10^4$.
	
	We put  qualitative
	characteristics of process with  diquark  and ways to  distinguish it from quark scattering  in model-independent way.
\end{abstract}
\newpage

\section{Introduction}

In the next 10 years there will be under operation new generation of experimental set ups as PANDA at FAIR (Germany)\cite{PANDA_Report:2020}, SPD at NICA (Russia) \cite{SPDproto:2021hnm} and J-PARC HI \cite{J-PARC-HI_White Paper:2019}. The unique features of these set ups contains such a peculiarities as
\begin{itemize}
\item full solid angle  $\Delta\Omega \sim4\pi$;
\item all type particle registration;
\item  luminosity  $\geq 10^{30}$ cm$^{-2}$ s$^{-1}$;
\item particle identification close to full energy range with high momentum resolution;
\item running  with antiproton, polarized and unpolarized light nuclear (deuterium) beams.
\end{itemize}

These new characteristics of experiments allow us to elaborate
new  research program in the energy range of $\sqrt{s_{NN}}$ from 3~GeV to 7~GeV.  At this energy range very interesting  high $p_T$ ($p_T > 0.5~GeV/c$) phenomena have been discovered a long ago but lacks sufficient theoretical and experimental thorough investigation. Some of them, as  enhancement of the high $p_T$ baryon matter production in exclusive scattering process, could be crucial  to receive proof of diquark constituent inside proton  and define diquark's properties  \cite{RNP:2005hw}.
 Another task could be to expand experimental data (investigate full set of the isotopic nucleon-nucleon interactions) about the  spin effects in the high $p_T$ region and clarify the nature of these effects  \cite{PolProg:2009hw
 }.

The best way to examine the nature of these phenomena is to research   exclusive reactions at the energy range of planned machines \cite{PANDA_Report:2020,SPDproto:2021hnm,J-PARC-HI_White Paper:2019}.
At  the higher energy range it is possible  to provide investigations in semi-exclusive and inclusive reactions and as a consequence such an information about final state as individual momentum  and type of scattered particles process will be lost.

The hypothesis of diquark as a constituent of barion brings new   theoretical  approaches to explain some experimental phenomena. They were introduced in baryon
spectroscopy shortly after  classification of meson resonances based on quark-antiquark system \cite{Ida},\cite{Lichtenberg:1967zz}. Later diquark bound states were used for theoretical prediction in various aspects of physics, from nuclear physics \cite{Suzuki:1989na} to models  of high-energy behaviour of meson and baryon cross-sections \cite{Martin:1986rtr}.

Nowadays there exist no universally accepted experimental evidence for the diquark states, but the conception of diquark are widely used   to describe the multiquark exotics, that was started with the discovery of X(3872) by the Belle Collaboration \cite{Belle:2003nnu} and prolonged by discovery of  series
of exotic states \cite{LHCb:2015yax},\cite{LHCb:2019kea}.

At that paper we want to discuss the possible role and ways of detection of diquarks in exclusive proton-proton  hard scattering process. There exist a lot of different  diquark models of baryons, in frames of which mass spectrum and form factors have been calculated \cite{Kroll:1990hg}, \cite{Kroll:1988fc}, \cite{Jaffe:1976ih},  \cite{Santopinto:2004hw}, but here we  distance  from model-dependent picture of diquark constituent of proton. Main goal of our paper is to find special processes and kinematics, where  contribution of diquark states could be enhanced and distinguished from quark contribution to the scattering process.

\section{Qualitative estimation}


Multiple interactions between constituents of different hadrons was considered in series of papers of  P.~Landshoff \cite{Landshoff} where the asymptotic contribution from disconnected graphs was considered. It was shown that scattered at large angle hadrons with successive independent scattering of its  constituents on one off the other has the scattering amplitude
$M_n\approx (\sqrt{s})^{4-n}$. Here $\sqrt s$ stands for initial energy of initial particles in center of mass frame.

In the well-known paper of S.~Brodsky and G.~Farrar \cite{Brodsky:1974vy} devoted to the estimation of the energy dependence of high-energy scattering process more accurate asymptotic behaviour of fixed angle scattering amplitude was calculated: $M_n\approx (\sqrt{s})^{4-n} (\sqrt s)^{L-1}$, where $L$ is the number of constituent pairs from different hadrons that have a interaction. Based on dimensional scaling laws
 the similar result was obtained in the paper of V.~A.~Matveev et al. \cite{Matveev:1972gb}, and the next scaling law should be observed  at large $s,t$ for scattering process $AB\to CD$:
\begin{gather}
\frac{d \sigma}{d t}(AB\to CD)\approx s^{1-n+L}f(\frac{s}{t}),
\label{eq:1}
\end{gather}
here $n$ is the total number of  constituents (fields) in particles $A,B,C,D$.

Accepting the QCD picture   for elastic $pp\to pp$ process  we require the scattering three quarks on three quarks  and as a consequence  could establish that  scattering energy dependence has to be $s^{-10}$. The direct measurement yields $d\sigma\approx s^{-9.7\pm0.3}$.
\begin{figure}[hbt]
	\centering
	\includegraphics[width=0.6\linewidth]{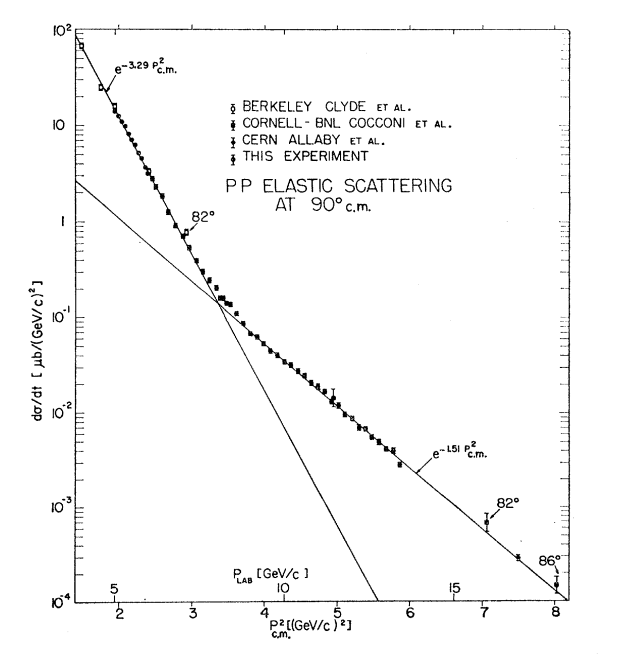}
	\caption {Plot of differential cross-section for large-angle  proton-proton elastic scattering. The picture was  taken from  \cite{Akerlof:1967zz}.}
	\label{Fig:10}
\end{figure}
Alternative schemes of quark-diquark proton's picture was known long ago \cite{Anselmino:1990wq}, and could give us the improvements at intermediate energies, where diquark states express some non-perturbative effect,  but straight prediction from eq. (\ref{eq:1}) gives us the energy dependence $s^{-6}$, which drastically differs from experiment. So we could conclude that possible contribution from diquark is  rather small compared to the quark-quark scattering, namely its contribution to the parton distribution function of proton is small . But in energy range about $\sqrt{s_{NN}}= 4~GeV$ there is some
deviation (about factor 2) from counting rules (\ref{eq:1}),  see Fig. \ref{Fig:10}. This irregularity may be
 phenomenological indication for  baryons as composed of a constituent diquark and quark.

To extract possible diquark scattering in proton's collision we suggest to consider  processes, where usual quark-quark scattering can be artificially  suppressed by phase volume restriction. It can be realized in $pp$ scattering with two additional particle (pion) creation, where special kinematic restrictions over final state are imposed.

\subsection{Final state kinematic restrictions}

\begin{figure}[hbt]
	\centering
	\includegraphics[width=0.3\linewidth]{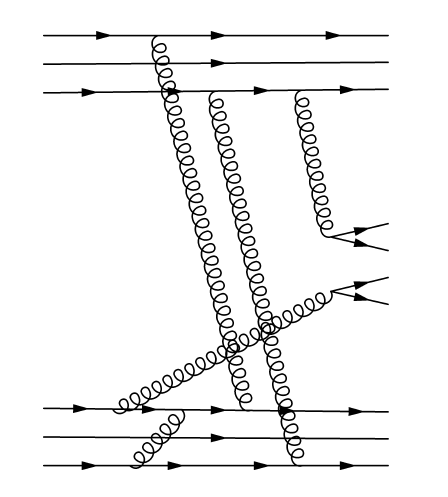}
	\caption {Example of diagram in $pp\to pp\pi\pi$ process where three quarks scatter on three quarks with two additional quark pair creation.}
	\label{Fig:3}
\end{figure}

In frames of QCD-parton scattering model for $pp\to pp\pi\pi$ process we should consider the scattering three quarks on three quarks with two additional quark pair creation (see Fig.\ref{Fig:3})
%
and further fragmentation to pair of proton and two pions where all particles have transverse momenta $p_T > 0.5$~ GeV/c.  If we eliminate  final states from  intermediate resonant decay as $\Delta^{+}$ we should not measure any correlation between scattering angles of proton and pion and similar rate of $\pi^\pm$ and $\pi^0$  pair production (see Sect. \ref{hadr} ).

\begin{figure}[hbt]
	\centering
	\includegraphics[width=0.3\linewidth]{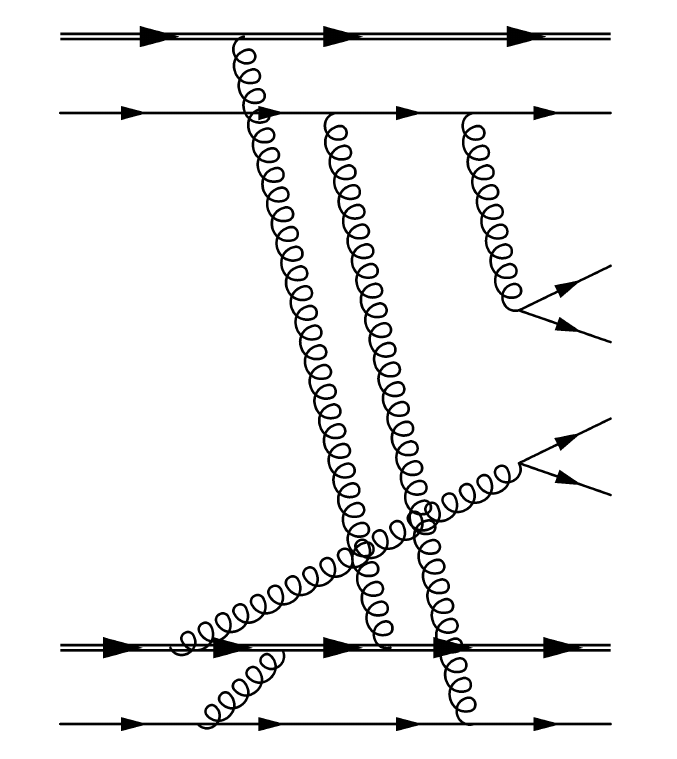}
	\caption {Example of diagram in $pp\to pp\pi\pi$ process where proton's partons (quark and diquark)scatters on each other  with two additional quark pair creation.}
	\label{Fig:4}
\end{figure}

Let us adopt the picture where proton consists of only two partons- quark and diquark. At that case we can consider $pp$ collision as  scattering of two types of partons on each other (Fig. \ref{Fig:4}), where scattered diquarks after hadronization process transforms to  protons and quarks to pion pair.

\begin{figure}[hbt]
	\centering
	\includegraphics[width=0.7\linewidth]{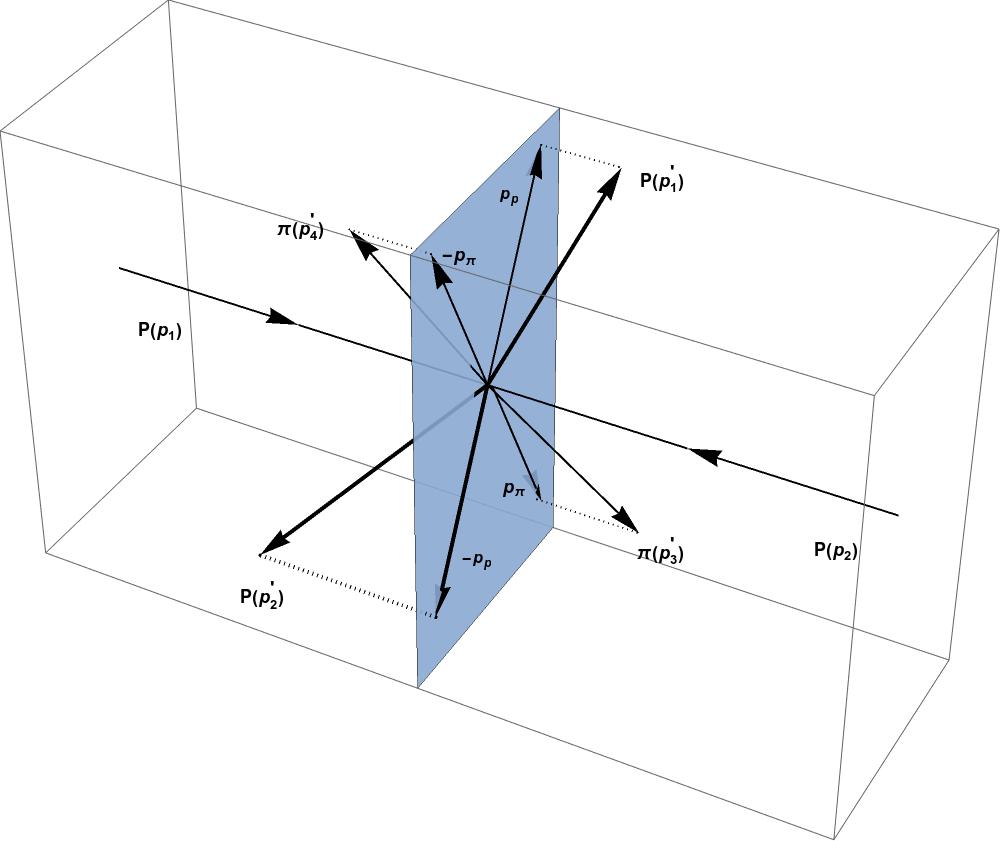}
	\caption {Kinematics of particles in $pp$ collision in the case of diquark-diquark scattering.}
	\label{Fig:2}
\end{figure}

Key feature of such a scattering in c.m.s. is the back-to-back scattering, where all scattered partons have equal and opposite impulses. Due to that incident   partons have different fraction of initial proton momentum,  only sum of transverse components of scattered parton's moments will be zero, namely, in $pp\to pp \pi\pi$ transverse momenta of final protons $p_p$ and pions $p_\pi$ are mutually cancels (see Fig. \ref{Fig:2}).
Taking into account that remarkable kinematic feature of $pp\to pp\pi\pi$  process in quark-diquark scattering picture we could impose restriction on possible final states,and consider  only impulses with sum of transverse momentum are less than detector angular resolution:
\begin{gather}
\frac{\Delta p_p}{p_p}\approx\frac{\Delta p_\pi}{p_\pi}\approx\delta.
\end{gather}
For example, angular momentum resolution of SPD detector on NICA is about \cite{SPDproto:2021hnm} $\delta\approx 0.01$. Neglecting possible statistics issues we could suppress usual quark-quark scattering contribution to $pp\to pp\pi\pi$ process due to final phase volume restrictions  up to $\delta^2\approx 10^{-4}$ and drastically improve
susceptibility of experiment to diquark contribution in scattering process.

\subsection{Hadronization}
\label{hadr}


Under well-known Lund model  of string fragmentation \cite{Artru:1979ye,Leader}, where the break-up of the string is attributed to quantum tunnelling with a probability  $-exp(-\pi m_T^2/\kappa)$, and $m_T$ is the quark pair transverse mass, we suggest that probability of hadronization of diquark to boson state dominates and we can safely  neglect the contrition from quark hadronization to proton. That assumption and requirement of heaving at final state only two protons and pion's pair lead us to the asymmetry in the rate of $\pi^\pm$ and $\pi^0$ pair production.

We take into account here only two  lowest quark flavours- $u$ and $d$, and neglect possible neutron formation from diquark due to its hard detection. Moreover, matrix element  of hard process does not depend over flavour in frames of QCD, so for ratio of pion production   we  consider  three possible configuration of partons at  final state, two symmetrical and one anti-symmetrical over flavour content of diquark and calculate total number of possible fragmentation channels:
\begin{itemize}
	\item we have two $(u,u)$ diquarks and two $d$ quarks from protons and two pair of quarks.
	As we consider fragmentation of diquark only to proton, here we have two channel for $\pi^\pm$ production and four for $\pi^0$ pair.
	\item we have  two $(u,d)$ diquarks and two $u$ quarks from protons and two pair of quarks.
	As in the case above,  we have two channel for $\pi^\pm$ production and four for $\pi^0$ pair.
	\item anti-symmetrical case, where there are $(u,u)$ and  $(u,d)$ diquarks and  $u$, $d$ quarks from protons and two pair of quarks. Here we have three channel for $\pi^\pm$ production and four for $\pi^0$ pair.
	\label{possSates}
\end{itemize}
Based on the assumption we have similar  parton distribution factors for diquarks $ud$ and $uu$,  and assuming only three possible  possible configuration of partons at  final state  we can conclude that production of  $\pi^0$ pairs  will dominate:
\begin{gather}
\frac{d\sigma(pp\to pp \pi^0\pi^0)}{d\sigma(pp\to pp\pi^+\pi^-)}\approx \frac{12}{7}\approx 1.7.
\label{eq:12}
\end{gather}

In the case when  $ud$ constituent  with flavour singlet  state totally  dominates over other diquark states inside proton (flavour triplet), we  have only one possible channel for $\pi^0$ formation, and  obtain dominant contribution of $\pi^0$ pairs:
\begin{gather}
\frac{d\sigma(pp\to pp \pi^+\pi^-)}{d\sigma(pp\to pp\pi^0\pi^0)}\approx 0.
\label{eq:13}
\end{gather}


\subsection{Energy dependence}


As we know form papers
\cite{Brodsky:1974vy,Matveev:1972gb}, the eq. (\ref{eq:1}) could be extended for the case of multi-particle exclusive large angle scattering. Based on the dimension of connected invariant amplitude we can conclude that $M_{n_i\to n_f}\approx s^{-\frac{1}{2}(n_i+n_f)+2}$, where $n_i,n_f$ is the number of quantum fields in initial and final state. Then it follows,using $\sigma\approx s^{-2}|M|^2$ that at multi-particle final state the eq. (\ref{eq:1}) modifies to
\begin{gather}
\frac{d \sigma}{d t_1 ...d p_ip_j ...}(AB\to CD...)\approx s^{2-n_i-n_f}f(\frac{p_i p_j}{s}).
\end{gather}
If we assume that there is no non-perturbative binding between quarks inside the hadrons, we obtain   the energy asymptotic behaviour of high energy scattering process at fixed ratios $ p_ip_j/s$ is proportional to $s^{2-3\times2-3\times2-2\times2}=s^{-14}$.

If we take into account only two constituents of proton, namely quark and diquark, the energy dependence of the process $pp\to pp\pi\pi$ will be $s^{2-2\times2-2\times2-2\times2}=s^{-10}$.

At the experiment we expect that any deviation from the energy asymptotic behaviour $s^{-14}$ will indicate on non-zero diquark contribution to process of $pp$ scattering. Moreover, as contribution of quark-quark scattering falls $s^{-4}$ faster than diquark one, we expect that energy dependence  with increasing $s$ will tends to $s^{-10}$. Such a picture will be clear evidence of non-zero diquark contribution to parton distribution functions of proton.


\section{Conclusion}

In the paper we describe possible scattering process, where contribution from diquark could be enhanced compared to ordinary quark scattering. We consider process of proton-proton scattering with additional pion par creation. Working within Lund model of string fragmentation, we expect dominant probability of diquark hadronization to boson. Keeping in mind that we have in final state only two protons  and pion pair, we establish that if diquark inside  proton  prevails in scattering process, we would have asymmetry in $\pi^0\pi^0$, $\pi^+ \pi^-$ pair creation rate (\ref{eq:12},\ref{eq:13}).

 Second criteria for discrimination between processes with and without diquarks is based on different  energy dependence of high-energy scattering process. Basing on dimensional scaling rules of matrix element we suppose  that contribution due to quark-quark scattering falls $s^{-4}$ faster than diquark one and energy dependence of the process could be used as distinction rule.

From the experimental data we know, that diquark distribution function of proton  is small and under the quark constituent picture we shouldn't be  relation between direction and magnitude  of final protons and pions. In the case of  diquark-quark or diquark-diquark  scattering  sum of transverse components of scattered parton's moments should be zero and transverse momenta of final protons $p_p$ and pions $p_\pi$ are mutually cancels. This feature helps one to select events with diquark scattering and  improve susceptibility of experiment to diquark up to $10^4$ for SPD set-up at NICA.

\section{Acknowledgements}

The work of V.V.B. was supported in part by the Heisenberg-Landau Program.

\end{document}